\documentclass[showpacs,preprintnumbers,amsmath,amssymb,prd]{revtex4}

\usepackage{graphicx}
\usepackage{dcolumn}
\usepackage{bm}
\begin{document}
\title{${\bar\nu}_l$ induced pion production from nuclei at $\sim$1 GeV}
\author{M. Rafi Alam}
\author{S. Chauhan}
\author{M. Sajjad Athar}
\email{sajathar@gmail.com}
\affiliation{Department of Physics, Aligarh Muslim University, Aligarh - 202 002, India}
\author{S. K. Singh}
\affiliation{H. N. B. Garhwal University, Srinagar - 246 174, India}

\date{\today}

\begin{abstract}
We have studied charged current $\bar\nu_l$ induced one pion production from  $^{12}$C and $^{16}$O nuclear targets at MiniBooNE and atmospheric antineutrino energies. 
The calculations have been done for the incoherent pion production process as well as for the pions coming from the hyperons in the 
quasielastic production of $\Lambda$ and $\Sigma$. The
calculations are done in the local density approximation. For the
inelastic processes the calculations have been done in the $\Delta$ dominance model and we take into account
the effect of Pauli blocking, Fermi motion of the nucleon and renormalization of $\Delta$ properties in
the nuclear medium. The effect of final state interaction(FSI) of pions is also taken into account. For the hyperon production, 
the nuclear medium effects due to Fermi motion and FSI effects due to hyperon-nucleon scattering have been taken into account. 
These results may be quite useful in the analysis of SciBooNE, MicroBooNE, MINER$\nu$A, and ArgoNeuT experiments when the pion analysis is done by using antineutrino beams.
\end{abstract}

\maketitle 

\section{Introduction}
Present and future neutrino oscillation experiments in the few-GeV energy region will focus on the measurement of
$CP$ violation in the neutrino sector, especially after observation of a nonzero 
value of $\theta_{13}$. This requires a precise determination of oscillation parameters to be determined by 
performing oscillation experiments with neutrino and antineutrino beams.
In order to facilitate extraction of various oscillation parameters from the analysis of these experiments,
it is desirable that the reaction cross sections for signal and background productions of leptons and pions 
in neutrino and antineutrino induced reactions from nuclear targets be known with high precision. 
In the intermediate energy region of a few-GeV, the cross sections for charged lepton production
are dominated by quasielastic and various inelastic processes involving pion production induced by charged 
and neutral currents in neutrino and antineutrino reactions from nuclear targets.

In general, antineutrino cross sections in the few-GeV energy region are not as well known experimentally as 
the neutrino cross sections, especially for the inelastic processes involving single
pion production. In particular there are no charged current(CC) antineutrino cross section measurements for 
pion production from nuclear targets in the energy region of $E_{\bar \nu} \le 1$  GeV. 
The only experimental data available for these processes in the low energy region of $1 < E_{\bar \nu} < 3$  
GeV are from the old CERN bubble chamber experiment~\cite{Bolognese:1979gf} using  propane mixed with 
freon where nuclear medium effects(NMEs) may be important. Indeed, a theoretical calculation done by Athar, Ahmad, and Singh ~\cite{Athar:2007wd} shows 
that NMEs are decisive in explaining the observed data. 

In the work of Athar, Ahmad, and Singh~\cite{Athar:2007wd}, the (CC) 
antineutrino induced pion production from nuclear targets was calculated in a $\Delta$ dominance model 
in which pions are produced through excitation of a $\Delta$ resonance. 
The $N-\Delta$ transition form factors for vector and axial vector currents are determined by using conserved vector current,
partially conserved axial current hypothesis and generalized Goldberger-Treiman relation for off-diagonal 
$N-\Delta$ transition~\cite{Fogli:1979cz}. Various NMEs on $\Delta$ production and 
their decay in a nuclear medium as well as the final state interaction(FSI) of pions were calculated in a model which 
has been earlier applied successfully to explain neutrino induced pion production in the energy region of 
$E_{\nu} \le 1.2$GeV~\cite{Singh:1998ha}.

In this paper, we apply the model as described in Refs.~\cite{Athar:2007wd,Singh:1998ha} 
with all the parameters fixed from the analysis of neutrino induced pion production from nuclear targets to study the
pion production induced by antineutrinos in nuclei. Except in the case of antineutrino reactions in 
nucleon and nuclei there is an additional contribution to the pion production from the quasielastic
$\Delta S=1$ processes in which hyperons like $\Lambda, \Sigma^{-,0}$ can be produced and decay subsequently to pions. 
These processes are generally Cabibbo suppressed as compared to the $\Delta$ production process
but could be important in the low energy region of antineutrinos relevant to MiniBooNE~\cite{AguilarArevalo:2008qa},
T2K~\cite{Abe:2011ks}, and atmospheric~\cite{Ashie:2005ik} neutrino experiments.
This possibility was discussed by Singh and Vacas~\cite{Singh:2006xp}, but no detailed quantitative estimates were made. 

In this study, we also include  the contribution of $\pi^-$ and $\pi^0$ which 
come  from the decay of hyperons($Y$) in the nuclear medium and present results for total production cross section of 
pions due to $\Delta$ and hyperon excitations in nuclei. The lepton energy as well as $Q^2$(square of the four momentum transfer) distributions are also 
presented and compared for the processes corresponding to $\Delta$ and
 $Y$ production leading to pions through their subsequent decay. In the case of hyperon production 
 NMEs in the production process as well as the FSI due to quasielastic and charge exchange hyperon nucleon scattering have also been 
 taken into account~\cite{Oset:1989ey}. However, NMEs on the hyperon 
 nucleon scattering processes are not taken into account. Moreover, we do not include the FSI of pions within the nuclear medium which are produced as a result of hyperon 
 decays. This is because the decay width of pionic decay modes of hyperons is highly suppressed in the 
 nuclear medium~\cite{Oset:1989ey}, making them live long enough to 
 pass through the nucleus and decay outside the nuclear medium and thus less affected by the strong 
 interaction of nuclear field. This is not the case with the pion produced through strong 
 decays of delta, which are further suppressed by the strong absorption of pions in the nuclear 
 medium~\cite{SajjadAthar:2009rd, VicenteVacas:1994sm}. Therefore, in the low energy region
 the Cabibbo suppression in the case of pion production through hyperons may be to some extent compensated by the 
 threshold suppression as well as strong pion absorption effects in the case of pions produced
  through delta excitation. The experimental results from SciBooNE~\cite{Cheng:2012yy}, MINER$\nu$A~\cite{Fields:2013zhk}, 
  MicroBooNE~\cite{Ignarra:2011yq}, and ArgoNeuT~\cite{Anderson:2011ce} in the antineutrino mode
  will be available in the near future, and the present results will be useful in the analysis of these experiments.
In Sec.~\ref{sec:pi_del}, we recall the formalism used for pion production through $\Delta$ 
excitation and, in Sec.~\ref{sec:pi_hyp}, the formalism used for production of pions through hyperon excitation.
 We present our results, conclusions, and discussions in Sec.~\ref{sec:res}.

\section{Pion production through $\Delta$ excitation}\label{sec:pi_del}
The basic reaction for the inelastic one pion production in nuclei, for an antineutrino interacting with a 
nucleon $N$ inside a nuclear target, is given by
\begin{eqnarray}\label{inelastic_reaction}
\bar\nu_{l}(k) + N(p) \rightarrow l^{+}(k^{\prime}) + \Delta(p_\Delta) & \nonumber \\
 {\rotatebox[origin=c]{180}{{\huge$\Lsh$}}} \quad \; &
\hspace*{-0.65cm} N^\prime (p^\prime) + \pi (k_\pi)
\end{eqnarray}
where $l=e, \mu$ and $N, N^\prime = p~ or~ n$. 
In the case of incoherent pion production in the $\Delta$ dominance model, the weak hadronic 
currents interacting with the nucleons in the nuclear medium excite a $\Delta$ resonance which decays into pions and nucleons. 
The pions interact with the nucleus inside the nuclear medium before coming out. 
The FSI of pions leading to elastic, charge exchange scattering, 
and the absorption of pions lead to reduction of pion yield. 
NMEs on $\Delta$ properties lead to modification in its mass and width which have 
been discussed earlier by Oset $et~ al.$~\cite{Oset:1987re, GarciaRecio:1989xa} and
 applied to explain the photon and electron induced pion production processes from nuclei. 

In the local density approximation, the expression for the total cross section
for the charged current one pion production is written as~\cite{Athar:2007wd}
\begin{eqnarray}\label{sigma_inelastic}
\sigma &=& \frac{1}{(4\pi)^5}\int_{r_{min}}^{r_{max}}~\rho_{N}(r) d\vec r~\int_{Q^{2}_{min}}^{Q^{2}_{max}}dQ^{2} 
 \int^{{k^\prime}_{max}}_{{k^\prime}_{min}} dk^{\prime}\int_{-1}^{+1}d\cos \theta_{\pi }\int_{0}^{2\pi}
d\phi_{\pi}~~\frac{\pi|\vec  k^{\prime}||\vec k_{\pi}|}{M E_{\nu}^2 E_{l}} \nonumber \\
&& \times \; \frac{1}{E_{p}^{\prime}+E_{\pi}\left(1-\frac{|\vec q|}{|\vec k_{\pi}|}
\cos\theta_{\pi }\right)}{\bar \sum} \sum|\mathcal M_{fi}|^2,
\end{eqnarray}
where $\rho_{N}(r)$ is the nucleon density 
defined in terms of nuclear density $\rho(r)$, for example, in the case of $\pi^-$ production, $\rho_{N}(r)=~\rho_{n}(r)~+~\frac{1}{9}\rho_{p}(r)$.

The transition matrix element $\mathcal M_{fi}$ is given by
\begin{equation}\label{matrix_element}
\mathcal M_{fi}=\sqrt{3}\frac{G_F}{\sqrt{2}}
\frac{f_{\pi N \Delta}}{m_{\pi}} cos\theta_{c} \bar u(p^{\prime}) k^{\sigma}_{\pi} {\mathcal P}_{\sigma \lambda} 
\mathcal O^{\lambda \alpha} u(p) L_{\alpha},
\end{equation}
where $L_{\alpha}$ is the leptonic current defined by 
\begin{equation}\label{lep_curr}
L_{\mu}=\bar{v_l}(k^\prime)\gamma_\mu(1-\gamma_5)v_\nu(k)
\end{equation}
and $\mathcal O^{\beta \alpha}$ is the $N-\Delta$ transition operator taken from Lalakulich, Paschos, and Piranishvili~\cite{Lalakulich:2006sw}.
${\mathcal P}^{\sigma \lambda}$ is the $\Delta$ propagator in momentum space and is given as 
\begin{equation}\label{width}
{\mathcal P}^{\sigma \lambda}=\frac{{\it P}^{\sigma \lambda}}{P^2-M_\Delta^2+iM_\Delta\Gamma}
\end{equation}
where ${\it P}^{\sigma \lambda}$ is the spin-3/2 projection operator and the delta decay width $\Gamma$ 
is taken to be an energy dependent $P-wave$ decay width~\cite{Oset:1987re, GarciaRecio:1989xa}:
\begin{equation}\label{Width}
\Gamma(W)=\frac{1}{6 \pi}\left(\frac{f_{\pi N \Delta}}{m_{\pi}}\right)^2 \frac{M}{W} |\vec q_{cm}|^3.
\end{equation}
$|\vec q_{cm}|$ is the pion momentum in the rest frame of the resonance, and $W$ is the center of mass energy. $f_{\pi N \Delta}$ is the $\pi N \Delta$ coupling constant taken as 2.12 for numerical calculations.
Inside the nuclear medium, the mass and width of delta are modified, which in the present calculation are 
taken into account following the works of 
Oset $et~ al.$~\cite{Oset:1987re, GarciaRecio:1989xa} and has been discussed 
in our earlier works~\cite{Athar:2007wd, SajjadAthar:2009rd}. The pions produced in this process are 
scattered and absorbed in the nuclear medium. 
This is treated in a Monte Carlo simulation which has been taken from Ref.~\cite{VicenteVacas:1994sm}.
\section{Pion production through hyperon excitation}\label{sec:pi_hyp}
$\Delta S=1$ hyperon $(Y)$ production processes induced by antineutrinos producing pions are given as Eq.(\ref{inelastic_reaction}) with $\Delta$ replaced by $Y$:
\begin{eqnarray}\label{hyp-rec}
{\bar{\nu_l}}(k) + N(p)\rightarrow l^+(k^\prime) + Y(p^\prime)&\nonumber \\
 {\rotatebox[origin=c]{180}{{\huge$\Lsh$}}} \quad \; &
\hspace*{-0.65cm} N^\prime (p^\prime) + \pi (k_\pi)
\end{eqnarray} 
where $Y=\Lambda, \Sigma^{0, -}$.
The differential scattering cross section for this process in the laboratory frame may be written as
\begin{eqnarray}
\label{crosv.eq}
d\sigma=\frac{\delta^4(k+p-k^\prime-p^\prime)}{16\pi^2 E_\nu M_N}\frac{d^3k^\prime}{2E_{k^\prime}}\frac{d^3p^\prime}{2E_{p^\prime}}\sum{\bar\sum}|{\cal{M}}|^2
\end{eqnarray}
where $M_N$ is the nucleon mass and ${\cal{M}}$ is the matrix 
element written as
\begin{equation}
{\cal{M}}=\frac{G_F}{\sqrt{2}}\sin\theta_c {\bar{v_l}}(k^\prime)\gamma^\mu(1 - \gamma^5)v_\nu(k)~J_\mu
\end{equation}
$J_\mu$ is the hadronic current, which is written as
\begin{eqnarray}\label{defff}
J_\mu&=&{\bar{u}_Y}(p^\prime)\left[\gamma_\mu f_1(q^2)+i\sigma_{\mu\nu}
\frac{q^\nu}{M+M_Y}f_2(q^2) - \gamma_\mu  \gamma_5 g_1(q^2) - \frac{q_\mu}{M_Y} \gamma_5 g_3(q^2) \right]u_N(p)
\end{eqnarray}
where $f_i(q^2)$ ($i=1,2$) and $g_i(q^2)$ ($i=1,3$) are the vector and axial vector $N-Y$($Y=\Lambda, \Sigma^{-}, \Sigma^{0}$) transition
form factors, respectively, which are given in terms of the functions $F^V_i(q^2)$ and
$D^V_i(q^2)$ corresponding to vector couplings and $F^A_i(q^2)$ and $D^A_i(q^2)$
corresponding to axial vector couplings and are taken from Ref.~\cite{Singh:2006xp}.

When the reactions shown in Eq.~(\ref{hyp-rec}) take place on nucleons which are bound in the
nucleus, Fermi motion and Pauli blocking
effects of initial nucleons are considered. The  Fermi 
motion effects are calculated in a local Fermi gas model, and the cross section is evaluated as a function 
of local Fermi momentum $p_F(r)$ and 
integrated over the whole nucleus. 
\begin{figure}
\includegraphics[height=0.4\textheight,width=0.7\textwidth]{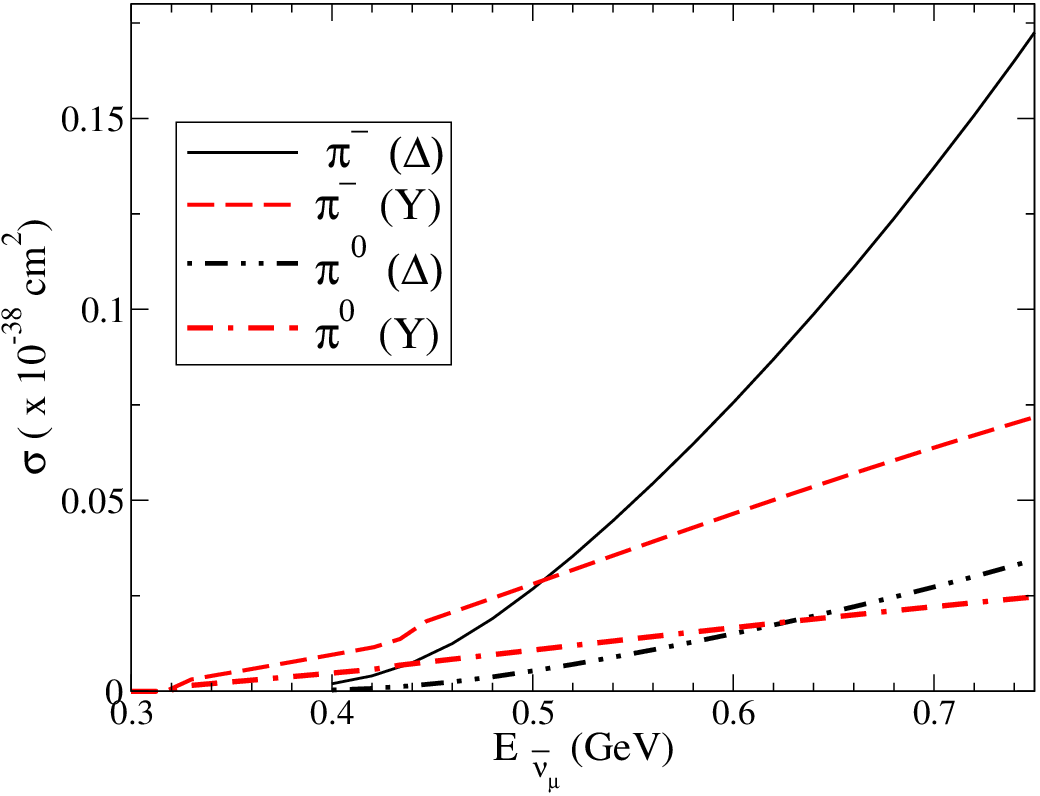}
\caption{$\sigma(E_{\bar\nu_\mu})$ vs $E_{\bar\nu_\mu}$ for $\pi^- \; \& \; \pi^0$ production in $^{12}$C, 
in the $\Delta$ dominance model and via intermediate hyperons.}
\label{pion.fig3}
\end{figure}
 The differential scattering cross section in this model is given by 
\begin{equation}
\frac{d\sigma}{dQ^{2} dE_l}=2{\int d^3r \int 
\frac{d^3p}{{(2\pi)}^3}n_N(p,r)\left[\frac{d\sigma}{dQ^{2} dE_l}\right]_{\rm{free}}},
\end{equation}
where $n_N(p,r)$ is the occupation number of the nucleon. $n_N(p,r)=1$ for $p\le p_{F_N}$ and is equal to zero 
for $p>p_{F_N}$, where $p_{F_N}$ is the Fermi momentum of the nucleon.
\begin{figure*}
\includegraphics[height=0.35\textheight,width=0.8\textwidth]{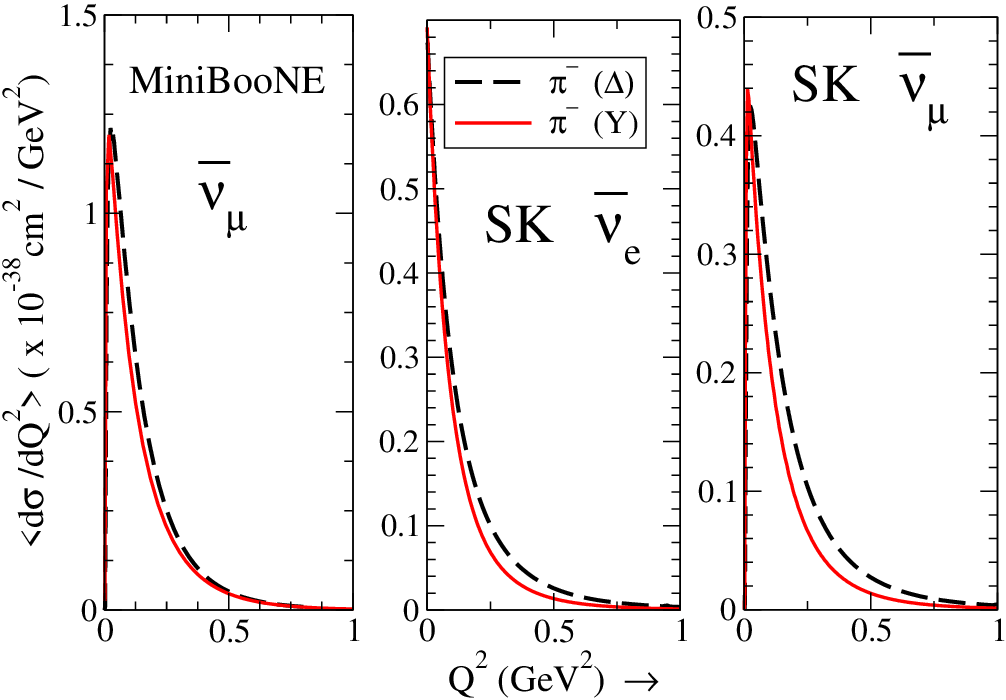}
\includegraphics[height=0.35\textheight,width=0.8\textwidth]{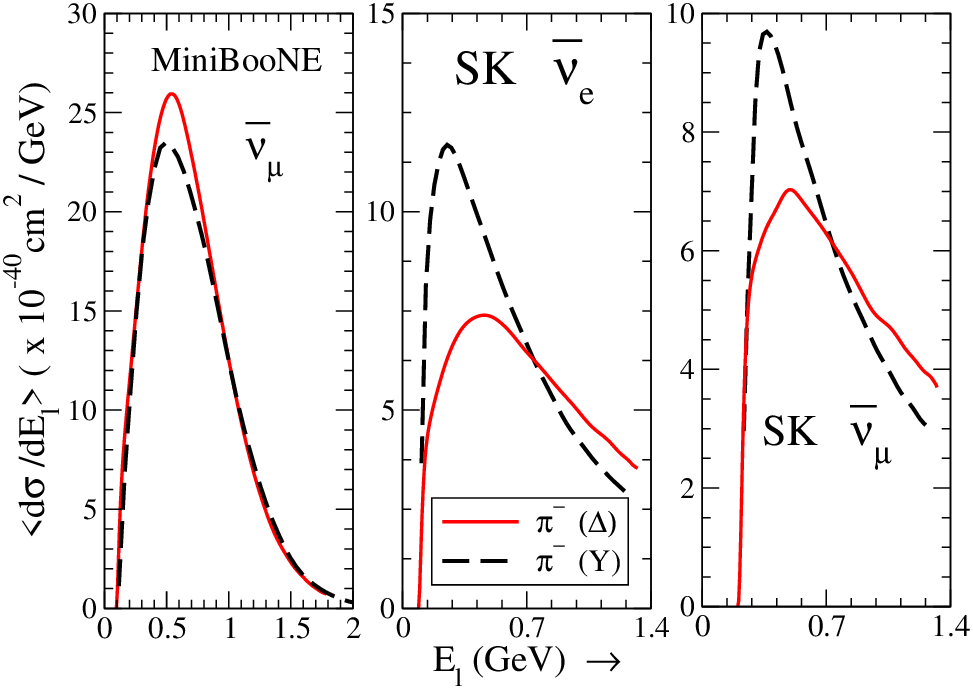}
\caption{$Q^2$ and lepton energy distributions for $\bar\nu_\mu$ induced reaction in $^{12}$C averaged over the MiniBooNE flux
and for $\bar\nu_{e,\mu}$ induced reaction in $^{16}$O averaged over the Super-Kamiokande flux are shown with nuclear medium and FSI effects. 
The $\pi^-$ production from hyperon excitations has been scaled by a
factor of 2.5.}
\label{difel.fig}
\end{figure*}
The produced hyperons are further affected by the FSI within 
the nucleus through the hyperon-nucleon quasielastic and
charge exchange scattering processes like $\Lambda + n \rightarrow \Sigma^- + p$, $\Lambda + n \rightarrow \Sigma^0 + n$, $\Sigma^- + p \rightarrow \Lambda + n$, 
$\Sigma^- + p \rightarrow \Sigma^0 + n$, etc. Because of such types of interaction in the nucleus, the probability of $\Lambda$ or $\Sigma$ production changes and has been
 taken into account by using the prescription given in Ref.~\cite{Singh:2006xp}. In this prescription an initial hyperon produced at a position 
$r$ within the nucleus interacts with a nucleon to produce a new hyperon state within a short distance $dl$ with a probability $P = P_{Y}dl$, where
$P_Y$ is the probability per unit length
 given by 
\[P_Y=\sigma_{Y+n \rightarrow f}(E)~\rho_{n}(r)~+~\sigma_{Y+p \rightarrow f}(E)~\rho_{p}(r),\]
 where $f$ denotes a possible final hyperon-nucleon [$Y_f(\Sigma ~{\text or}~ \Lambda) + N(n~or~p)$] state with energy $E$ in the hyperon-nucleon center of mass system, 
$\rho_{n}(r)[\rho_{p}(r)]$ is the local density of the neutron(proton) in the nucleus, and $\sigma$ is the total cross section for a charged current channel
 like $Y(\Sigma ~{\text or}~ \Lambda)~+~N(n~or~p) \rightarrow f$~\cite{Singh:2006xp}. Now a particular channel is selected, which gives rise to a hyperon $Y_f$ in the final state with the probability $P$. 
For the selected channel, the Pauli blocking effect is taken 
into account by first randomly selecting a nucleon in the local Fermi sea. Then a random scattering angle is generated in the hyperon-nucleon center of mass system assuming the cross sections to be isotropic.
 By using this information, hyperon and nucleon momenta are calculated and Lorentz boosted to the lab frame. 
 If the nucleon in the final state has momenta above the Fermi momenta, we have a new hyperon
type($Y_f$) and/or a new direction and energy of the initial hyperon($Y_i$). This process is continued until the hyperon gets out of the nucleus.

 As the decay modes of hyperons to pions is highly suppressed in the 
 nuclear medium~\cite{Oset:1989ey}, making them live long enough to 
 pass through the nucleus and decay outside the nuclear medium, therefore, the produced pions are less affected by the strong 
 interaction of nuclear field, and their FSI have not been taken into account. 

\section{Results and Discussion}\label{sec:res}
We find that in the case of pions produced through $\Delta$ excitations NMEs lead to a reduction of around 32$\%$ - 40$\%$ for 
antineutrino energies 0.4 $< E_{\bar\nu_\mu} <$ 1.2GeV. When
pion absorption effects are taken into account, the total
reduction in the cross section is around 40$\%$ - 50$\%$. Furthermore,
the production cross sections of these pions are inhibited by the threshold effects, and there is almost no cross section until $E_{\bar{\nu}}$ = 400 MeV. 
In the case of pions produced through the hyperon production, there is an overall suppression by a factor of sin$^{2}\theta_{c}$ but
these are kinematically favored as the $\Lambda$ production starts around $E_{\bar{\nu}}$ = 250 MeV while $\Sigma^-$ and $\Sigma^0$
production starts around $E_{\bar{\nu}}$ = 325 MeV. We find that the effect of the 
Fermi motion of the initial nucleons is quite small 
on the quasielastic production of hyperons, and the effect of FSI results in an enhancement in the
$\Lambda$ production cross section and a suppression in the $\Sigma$
 production cross section. This is due to the fact that $\Sigma^{-,0}$ can disappear through the quasielastic 
 processes like $\Sigma^-+p\rightarrow\Lambda^0+n$ or 
$\Sigma^0+n\rightarrow\Lambda^0+n$ or through some other channels, whereas this is inhibited for the $\Lambda$ to 
change into $\Sigma$ due to the difference in masses. We find that the effect 
of the FSI increases with the increase in the mass number of the nucleus. 
\begin{figure*}
\includegraphics[height=0.35\textheight,width=0.8\textwidth]{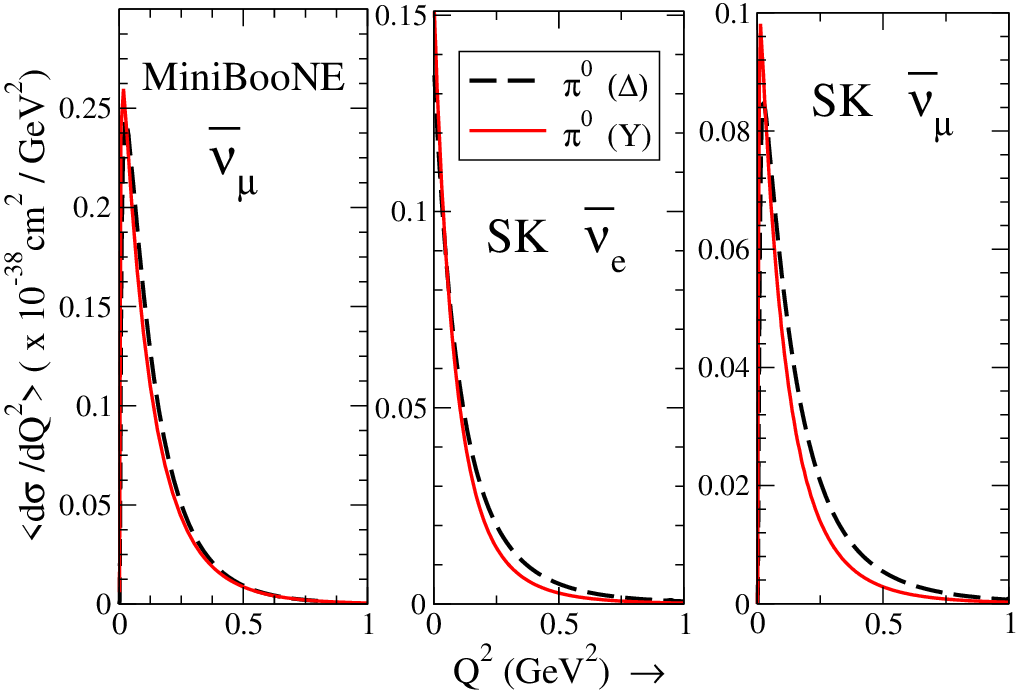}
\includegraphics[height=0.35\textheight,width=0.8\textwidth]{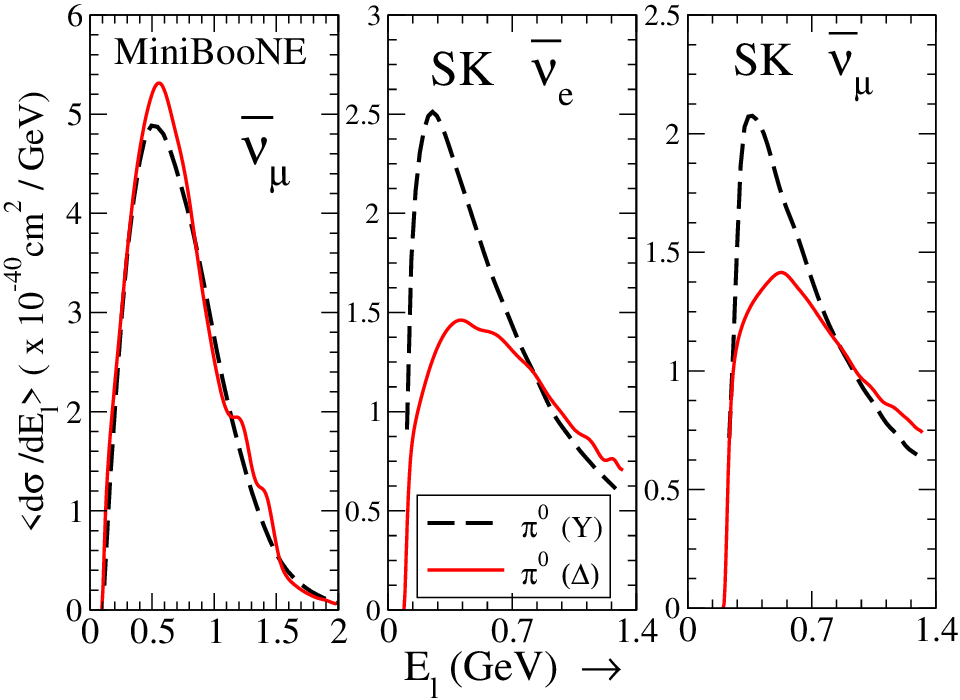}
\caption{$Q^2$ and lepton energy distributions for $\pi^0$ production. The lines have the same meaning as in Fig.~\ref{difel.fig}. 
The $\pi^0$ production from hyperon excitations has been scaled by a factor of 1.33.}
\label{difel1.fig}
\end{figure*}

In Fig.~\ref{pion.fig3}, we compare the  results for the cross sections $\sigma(E_{\bar{\nu}})$ 
for one pion production obtained from the hyperon excitation and the $\Delta$ excitation.
 These results are presented with nuclear medium and FSI effects for the 
 pions interacting with the residual nucleus for $\pi^-$ as well as $\pi^0$ productions obtained
  in the delta dominance model and the pions obtained from the hyperons where the medium 
  effects in hyperon production and FSI due to hyperon-nucleon interaction in the nuclear medium have been taken into account. 
 When the results for the cross sections are compared in these two processes, we find that, 
 at lower antineutrino energies $E_{\bar{\nu}} < $ 500 MeV, the contribution of $\pi^-$ from the hyperon excitation is 
 more than the pions coming from the $\Delta$ excitation. For $E_{\bar{\nu}} > $ 500 MeV the contribution from $\Delta$ excitation starts dominating and at $E_{\bar{\nu}}$ = 1 GeV the 
  contribution of $\pi^-$ from the hyperon excitation is around 20$\%$ of the total $\pi^-$ production. 
  In the case of $\bar\nu_l$ induced $\pi^0$ production, the contribution of $\pi^0$ from the hyperon excitation is larger up to antineutrino energy of 
  650 MeV than 
  the contribution of $\pi^0$ from the $\Delta$ excitation. At $E_{\bar{\nu}}$ = 1 GeV the 
  contribution of $\pi^0$ from the hyperon excitation is around 30$\%$ of the total $\pi^0$ production. Qualitatively similar effects are also obtained in the case of $^{16}$O which are in agreement 
with the results earlier obtained by Singh and Vacas~\cite{Singh:2006xp}.
  
Motivated by these results in  Figs.\ref{difel.fig} and \ref{difel1.fig},
we have presented, respectively, the
results for the $Q^2$ distribution and lepton energy distribution, averaged over the
MiniBooNE~\cite{AguilarArevalo:2013hm} and Super-Kamiokande~\cite{Honda:2006qj} spectra. We must point out that the different scales have been used for
$\pi^-$(scaled by 2.5) and $\pi^0$(scaled by 1.33) productions from hyperon excitations. The results are presented for the
differential cross sections calculated with nuclear medium and  pion absorption effects for the pions obtained 
from the decay of $\Delta$ excitation. 
For the $Q^2$ distribution of pions obtained from the hyperon excitation,
we find that when the contribution for the pions coming from all the hyperons mentioned in Eq.(\ref{hyp-rec}) 
 are taken together, there is negligible effect of nuclear medium and FSI on the total pion production. 
  We observe that, in the peak region of $Q^2$ distribution, the contribution of $\pi^-$ from the hyperon excitations is almost 40$\%$ to the
 contribution of $\pi^-$ from the $\Delta$ excitation, whereas in the case of $\pi^0$ production the contribution from hyperon excitations is almost 75$\%$ to the
 contribution of $\pi^0$ from the $\Delta$ excitation in the peak region. Thus in the case of $\pi^-$/$\pi^0$ the contribution
 from hyperon production is quite significant and must be taken into account in all the Monte Carlo
  generators looking for the pion events through the inelastic channel. 
  
  In the case of $E_l$ distribution of leptons accompanied by a $\pi^-$($\pi^0$), we find that in the peak region 
  the contribution from hyperon excitations is almost 40$\%$(75$\%$) of the
 contribution from the $\Delta$ excitation if the differential cross sections $\frac{d\sigma}{dE_l}$ are folded with MiniBooNE flux. When the lepton energy distribution $\frac{d\sigma}{dE_l}$ 
is folded with 
the atmospheric antineutrino flux, the contribution from hyperon excitations to the $\pi^-$ production is 
 around 45$\%$ of the $\pi^-$ from  $\Delta$ excitation,  whereas there is almost equal contribution of 
 $\pi^0$ from $\Delta$ and hyperon excitations. 
 
To summarize our results, we find that the reduction due to nuclear medium and FSI 
effects in the case of pions obtained from $\Delta$ excitation
is large enough to compensate for Cabibbo suppression of pions produced through hyperon excitations. Because of this, pion production from hyperons is larger than the pion production 
from $\Delta$ excitation up to the antineutrino energies of about 500 MeV for $\pi^-$ production and 650 MeV for $\pi^0$ production. This results in a 
significant contribution to pion production from hyperon excitation when folded with the spectra of
MiniBooNE or atmospheric antineutrino fluxes. These results will have important implications in the analysis of SciBooNE, MicroBooNE, MINER$\nu$A and
ArgoNeuT experiments when the data come up  
using antineutrino beams.
\section{Acknowledgements}
M. R. A. wants to thank Council of Scientific and Industrial Research(CSIR) for a senior research fellowship. S. C. is thankful to Department of Science
and Technology(DST) Women Scientist Scheme(WOS-A), Government of India for providing financial assistance under Grant No. SR/WOS-A/PS-10/2011. M. S. A. is thankful to Department of Science
and Technology(DST), Government of India for providing financial assistance under Grant No. SR/S2/HEP-18/2012.

\end{document}